\documentclass[fleqn,usenatbib,useAMS]{mnras}

\usepackage{graphicx}             
\usepackage{amsmath}              
\usepackage{upgreek}              
\usepackage{amssymb}              
\usepackage{bm}                   
\usepackage{CJKutf8}              
\usepackage{comment}              
\usepackage{tabularx}             
\usepackage{threeparttable}       
\usepackage{booktabs}             
\usepackage{soul}                 
\usepackage{ulem}                 
\usepackage{multirow}             
\usepackage[T1]{fontenc}          
\usepackage{makecell}
\usepackage[dvipsnames]{xcolor}   
\usepackage{microtype}            
\DisableLigatures[f]{encoding = *, family = *} 
\emergencystretch=\maxdimen       
\newcommand{\fracp}[2]{\frac{\partial{#1}}{\partial{#2}}}

\defcitealias{Li2022}{Paper I}

\usepackage{ae,aecompl}
\usepackage{multicol}        
\usepackage{newtxtext,newtxmath}

\title[Dust Growth near Truncation Radius. II.]{Dust Accumulation near the Magnetospheric Truncation of Protoplanetary Discs. II. The Effects of Opacity and Thermal Evolution}

\author[R. Li, Y. -X. Chen, and D. N. C. Lin]{
Rixin Li$^{1}$
\begin{CJK*}{UTF8}{gbsn}
  (李日新)
\end{CJK*} \thanks{Contact e-mail: \href{mailto:rixin@berkeley.edu}{rixin@berkeley.edu}}\thanks{51 Pegasi b Fellow},
Yi-Xian Chen$^{2}$
\begin{CJK*}{UTF8}{gbsn}
  (陈逸贤)
\end{CJK*}
, and Douglas N. C. Lin$^{3,4}$
\begin{CJK*}{UTF8}{gbsn}
  (林潮)
\end{CJK*}
\\
$^{1}$Department of Astronomy, University of California, Berkeley, Berkeley, CA 94720, USA \\
$^{2}$Department of Astrophysical Sciences, Princeton University, Princeton, NJ 08544, USA \\
$^{3}$Department of Astronomy, University of California, Santa Cruz, CA 95064, USA \\
$^{4}$Institute for Advanced Studies, Tsinghua University, Beijing, 100084, People’s Republic of China
}

\date{Accepted 2024 February 19. Received 2024 February 15; in original form 2023 September 30}
\pubyear{2024}
\begin{document}
\label{firstpage}
\pagerange{\pageref{firstpage}--\pageref{lastpage}}
\maketitle

\begin{abstract}
  Dust trapping in the global pressure bump induced by magnetospheric truncation offers a promising formation mechanism for close-in super-Earths/sub-Neptunes.  These planets likely form in evolved protoplanetary discs, where the gas temperature at the expanding truncation radius become amiable to refractory solids.  However, dust accumulation may alter the disc opacity such that thermal evolution is inevitable.  To better understand how thermodynamics affects this planet formation pathway, we conduct a suite of local dust evolution simulations in an idealized inner disc model.  Our calculations take into account self-consistent opacity-dependent temperature changes as well as dust evaporation and vapour condensation.  We find that disc thermal evolution regulates dust growth and evolution, discouraging any accumulation of small particles that drives the increase of opacity and temperature.  Significant retention of dust mass takes place when the disc environments allow runaway growth of large solids beyond the fragmentation barrier, where small particles are then swept up and preserved.  Our results further validate dust accumulation near disc truncation as a promising mechanism to form close-in planets.
\end{abstract}


\section{Introduction}
\label{sec:intro}

Short-period super-Earths (also referred to as sub-Neptunes or Kepler planets) are the most common type of currently observable exoplanets \citep[e.g.,][]{Howard2010, Dressing2015}. However, their distinct properties (e.g., tight orbital configurations, insensitivity to host metallicities, mass-radius valley) and the lack of their analogues in the Solar System challenge the conventional planet formation theory \citep{Petigura2017,Johnson2017,Fulton2017,Petigura2018,Zhu2018,Kunimoto2020,Reiners2022,Stefansson2023}.

Several formation channels for such super-Earths have been proposed over the years, with {a broad} consensus that super-Earths are most likely formed out of materials originated from the external disc (i.e., \textit{ex situ} formation) ({\citealt{Morbidelli2016}, see also their Fig. 5 and references therein}; \citealt{Izidoro2021}).  The remaining key question is when and where growth takes place during inward migration of solids.  {If the initial growth of planetary embryos is efficient, they may form in the external disc and then migrate into the tight orbits \citep[e.g.,][]{Ida2010, Kley2012}.}  Pebble accretion may help embryos grow much further during and after migration \citep[e.g.,][]{Lambrecht2014, Bitsch2015, Johansen2017, Lambrechts2019}.

Alternatively, rapid inward drift of dust into traps induced by pressure bumps at short-period orbits may retain solid mass fast enough to form embryos and supply their further growth.  The dead zone inner boundary (DZIB) may serve this purpose, but recent disc modeling suggested that the DZIB either locates far out or may not induce a pressure bump at all \citep[e.g.,][]{Chatterjee2014, HuXiao2016, HuXiao2018, Jankovic2021, Jankovic2022}.

In \citet[][hereafter \citetalias{Li2022}]{Li2022}, we proposed and demonstrated that the {global pressure maxima} induced by magnetospheric truncation in evolved discs may trap dust efficiently, providing a promising formation pathway for close-in super-Earths.  Specifically, two sub-pathways for effective dust retention operate at different parameter regimes.  When the turbulent relative velocity between dust (correlated with the Shakura–Sunyaev $\alpha$-parameter \citep{Shakura1973} and the gas sound speed $c_{\rm s}$) is relatively low and the threshold velocity for fragmentation $u_{\rm f}$ is high (e.g., due to the sticky surface layer on dust at a high temperature), or when large solids can be rapidly delivered from the outer disc, a fraction of particles may grow fast and surpass the fragmentation barrier.  This process triggers an accelerated accumulation of progressively larger particles, {eventually reaching the typical size of planetesimals (referred to as the \textit{breakthrough scenario})}.  Conversely, in a different scenario characterized by a high turbulent relative velocity and a low threshold velocity for fragmentation, coagulation growth beyond small grains is highly suppressed. However, when the dust supply from the outer disc is substantial enough to counteract the removal through funnel flows, the overall dust mass can still gradually rise, eventually reaching a point where gravitational instability (GI) produces planetesimals (referred to as the \textit{feedback+GI scenario}).

Dust retention at the truncation radius $R_{\rm T}$, which converges with the corotation radius $R_{\rm Co}$ at low accretion rates (e.g., \citealt{Long2005}; \citealt{Bouvier2007}; \citealt{Romanova2008}; \citealt{Thanathibodee2023}; \citealt{Zhu2023}; Pittman et al. in prep.), is of great astrophysical interest by itself even if the accumulated dust does not end up in large terrestrial planets. Dust features at these location are one of the favourable explanations proposed for ``dipper'' stars \citep[e.g.,][]{Bouvier1999,Cody2014,Ansdell2016a,Stauffer2017,Robinson2021,Capistrant2022}.  Moreover, small rocky bodies formed from the accumulated dust near $R_{\rm Co}$ may serve as sources of materials that produce complex periodic variables, a new class of variable star \citep[][]{Bouma2023}.

In \citetalias{Li2022}, we adopted an idealized model with a static gas disc for simplicity, which allowed us to focus on a preliminary exploration of dust evolution.  In this work, we take into account the dynamical thermal evolution of the gas disc caused by dust accumulation and the resulting changes in dust opacity.  Our goal is to understand how opacity-driven thermal evolution affects dust evolution as well as this new planet formation pathway.

The paper is organized as follows.  Section \ref{sec:coag_model} first describes our dust evolution model, including our new considerations on dust opacity and disc thermal evolution.  Section \ref{sec:results} then presents our simulation results.  Finally, Section \ref{sec:final} summarizes our key findings and discusses the important implications. 

\section{Dust Evolution Model with Opacity and evaporation}
\label{sec:coag_model}

We use the implicit dust coagulation-fragmentation code \texttt{Rubble} \citepalias{Li2022} to model the evolution of dust size distribution, dust surface density, and the thermal evolution between dust and vapour at the dust accumulation radius $R_{\rm accu}$ (i.e. the truncation-induced global pressure maximum) , in a radiative disc model initialized based on \citet{Ali-Dib2020}.  In this work, \texttt{Rubble} has been upgraded to run on graphics processing units (GPUs) by utilizing \texttt{PyTorch} \citep{PyTorch}, which greatly speeds up our simulations.

In this section, we briefly reiterate our base numerical scheme to model dust size evolution.  Sections \ref{subsec:opacity} and \ref{subsec:ec} then describe how \texttt{Rubble} calculates dust opacity and solid evaporation and condensation due to temperature changes.  Finally, Section \ref{subsec:setup} details the parameters for our simulations. 

\subsection{Basic Scheme}
\label{subsec:basic_scheme}
\texttt{Rubble} solves the Smoluchowski equation
\begin{equation}\label{eq:Seq}
    \fracp{}{t} N(m) = \int\int_0^\infty M(m, m', m'') N(m') N(m'') \textnormal{d}m' \textnormal{d}m'',
\end{equation}
where $N(m)\equiv \textnormal{d}N/\textnormal{d}m$ is the vertically integrated dust surface number density in a mass bin, $M(m, m', m'')$ is the full kernel that consists of coagulation kernel and fragmentation kernel, {which elucidates the likelihood of particles of masses $m'$ and $m''$ colliding with each other to produce target mass of $m$}.

To relate the total dust surface density $\Sigma_{\rm d}$ and $N(m)$, we define the vertically integrated dust surface density distribution per logarithmic bin of grain radius $\sigma(a)$, and

\begin{equation}
  \Sigma_{\rm d} = \int_0^{\infty} \sigma(a)\ \textnormal{d} \log a,
\end{equation}
where
\begin{equation}
  \sigma(a) = N(m) \cdot 3 m^2 = \frac{dN}{d\log m} 3 m,
\end{equation}
where $\textnormal{d}N/\textnormal{d}\log m$ is the vertically integrated dust surface number density in a logarithmic mass interval and is the quantity that our implicit code actually evolves.

\subsection{Opacity and Energy Equation}
\label{subsec:opacity}

We calculate the dust opacity from a full size distribution using the prescription from \citet{ChenYX2020}, modified from the single-species opacity formula of \citet{Ormel2014}.  For each size bin in the dust distribution with characteristic size $a_{k}$, the metallicity is $Z_{ k}=\sigma_{k}\textnormal{d}\log a / \Sigma_{\rm g}$ where $\Sigma_{\rm g}$ is the gas surface density.  The total opacity consisting of the dust and gas opacity is (in cgs units)
\begin{equation}
  \kappa=\kappa_{\mathrm{gas}}+\kappa_{\mathrm{dust}}=10^{-8} \rho_{\rm gas}^{2/3} T^{3}+\sum_{k} \kappa_{k} Q_{k},
\end{equation}
where $\rho_{\rm g}$ and $T$ denote the midplane gas density and temperature, $\kappa_{k}$ is the dust geometric opacity, and $Q_k$ is the efficiency coefficient.  The first term in the equation above is the gas opacity that only dominates at high temperature ($\gtrsim$ $2000$K or when the gas is dust-free), and the second term denotes the dust opacity that become dominant below $2000$K.  The gas opacity which dominates at large temperature is approximated by analytical expressions from \citet{BellLin1994}. It will be dominated by dust opacity below 2000K
\footnote{In principle, $\kappa_{\rm gas}$ may be non-negligible if the gas density and metallicity is very high, as in the case of planetary atmospheres \citep{Valencia2013,Freedman2014}.  However, it is never the case in the
context of the work presented here.}.
Since this opacity matters in the calculation of radiative energy flux of the gas profile (optical depth $\tau = \kappa \Sigma_{\rm g}/2$ ), we need to scale the dust cross section to that of per unit \textit{gas mass}.  Thus, 
\begin{equation}
    \kappa_{k}=\frac{3}{4 \rho_{s} a_{k}} Z_{k},\ \ \  Q_{k}=\min \left(0.3 x_{k}, 2\right),
\end{equation}
where $\rho_s$ is the dust internal density, and $x_{k}=2 \pi a_{k} / \lambda_{\max}$, where $\lambda_{\max } = 0.29\ \text{cm}/T$ is the peak wavelength of blackbody radiation from Wien’s law.

The disc thermal evolution is then driven by the following energy equation
\begin{align}
    \Sigma_{\rm g} C_v \dfrac{dT}{dt} 
    &=\dot{Q}_{+} - \dot{Q}_{-,\rm rad} -   \dot{Q}_{-,\rm d2v} \\
    &= \frac{9}{4} \Sigma_{\rm g} \nu \Omega^2 - \sigma_{\rm SB}\frac{T^4 \Sigma_{\rm g} \kappa}{(1 + \Sigma_{\rm g}^2 \kappa^2)}  - \dot{\Sigma}_{\rm d2v} L_0,
    \label{eq:energy}
\end{align}
where $C_v=1.7\times 10^{8}$ erg g$^{-1}$ K$^{-1}$ is the specific heat of hydrogen gas, $\dot{Q}_+$ is the viscous-heating source term \citep{Pringle1981} at the corotation radius $R_{\rm Co}$, where the the viscosity $\nu = \alpha c_{\rm s} H$ and $\Omega = 2 \pi /P_\star$, where $H$ is the gas scale height and $P_\star$ is the stellar spin period.  Ohmic heating near $R_{\rm Co}$ is negligible and the stellar irradiation is also weak in the inner disc region \citep{GaraudLin2007}.  The last two terms on the right hand side are the cooling source terms, where $\dot{Q}_{-,\rm rad} = 2\sigma_{\rm SB} T_{\rm s}^4$ is the radiative cooling term and $\dot{Q}_{-,\rm d2v}$ is the dust-to-vapour evaporation cooling term.  For the first cooling term, $\sigma_{\rm SB}$ is the Stefan–Boltzmann constant and $T_{\rm s}$ is the effective surface temperature, which is approximately related to the midplane temperature as
\begin{equation}
    {T^4} = T_{\rm s}^4 (\tau + \dfrac1{4\tau}),
\end{equation}
where $\tau$ is the optical depth and the formula above takes into account both optically thin and optically thick situations.  For the second cooling term, $\dot{Q}_{-,\rm d2v}$ is proportional to the dust-to-vapour mass conversion rate per unit area $\dot{\Sigma}_{\rm d2v}$, and $L_0$ is the latent heat needed during dust evaporation.

\begin{table*}
  \caption{Simulation Parameters \label{tab:paras}}
  \begin{tabular}{c|c|c|c|c|c|c}
  \hline
  Setup &
  $\alpha$ &
  $H$ &
  $T_{\rm rad,0}$ &
  $\Sigma_{\rm g}$ & 
  $\Sigma_{\rm v,0}$ & 
  $r_{\mathrm{d,critD}}$ \\
  &
  &
  [au] &
  [K] &
  [g cm$^{-2}$] &
  [g cm$^{-2}$] &
  [cm] \\
  \hline\hline
  A & 1.00e-3 & 0.00313 & 1880.60 & 1451.70 & 1.22e+02 & 1.47e-02 \\
  B & 1.78e-3 & 0.00296 & 1676.09 &  915.96 & 2.70e+00 & 1.56e-02 \\
  C & 3.16e-3 & 0.00279 & 1493.82 &  577.93 & 3.73e-02   & 1.65e-02 \\
  D & 5.62e-3 & 0.00264 & 1331.37 &  364.65 & 3.02e-04 & 1.75e-02 \\
  E & 1.00e-2 & 0.00249 & 1186.58 &  230.08 & 1.36e-06 & 1.86e-02 \\
  \hline\hline
  \multicolumn{7}{c}{} \\[-0.5em]
  \multicolumn{7}{c}{All available choices$^{*}$ of $u_{\rm f}$, $a_{\rm supp,max}$, and $Z_{\rm supp}$} \\[0.25em]
  \hline
  \multicolumn{3}{c|}{$u_{\rm f}$ [cm s$^{-1}$]} &
  \multicolumn{4}{c}{100, 178, 316, 562, 1000} \\
  \multicolumn{3}{c|}{$a_{\rm supp,max}$ [cm]} &
  \multicolumn{4}{c}{30, 100, 300, 1000} \\
  \multicolumn{3}{c|}{$Z_{\rm supp}$} &
  \multicolumn{4}{c}{0.01, 0.05} \\
  \hline
  \end{tabular} \\
  \begin{minipage}[]{0.75\textwidth}
    {\large N}OTE --- The simulation parameters (except $\Sigma_{\rm v,0}$) listed in this table are the same as those listed in Table 1 in \citetalias[][]{Li2022} (see Section 4.3).  For all setups, we assume the stellar radius $R_{\star}=1.8 R_\odot$, the strength of the stellar dipole magnetic ﬁeld $B_{\star}=1.0$ kG, the stellar spin period $P_{\star}$ = 8 days, the ratio between the truncation radius and the dust accumulation radius $f_{\rm out} = R_{\rm accu}/R_{\rm T} = 1.25$, the accretion rate $\dot{M} = 3.06$e$-9 M_\odot$ yr$^{-1}$, and $R_{\rm accu} = 0.098$ au.
  \end{minipage}
\end{table*}

\subsection{Evaporation and Condensation}
\label{subsec:ec}

To model the dust evolution that includes evaporation and condensation, we assume most solids are silica (e.g., SiO$_2$) and develop a semi-implicit scheme based on the following equations \citep{Schoonenberg2017}
\begin{align}
  \dot{\Sigma}_{\rm v} &= \dot{\Sigma}_{\rm d2v} - \dot{\Sigma}_{\rm v2d}, \label{eq:dot_Sigma_v} \\
  \dot{\Sigma}_{\rm d} &= \dot{\Sigma}_{\rm v2d} - \dot{\Sigma}_{\rm d2v}, \label{eq:dot_Sigma_d} 
\end{align}
where the dust-to-vapour conversion rate is $\dot{\Sigma}_{\rm d2v} = R_{\rm e}\Sigma_{\rm d}$ and the vapour-to-dust conversion rate is $\dot{\Sigma}_{\rm v2d} = R_{\rm c}\Sigma_{\rm v}\Sigma_{\rm d}$, where $\Sigma_{\rm v}$ is the vapour surface density and the coefficients
\begin{align}
  R_{\rm e} &= 8 \sqrt{2\pi} \frac{a^2}{m} \sqrt{\frac{\mu_{\rm SiO_2}}{k_{\rm B}T}} P_{\rm eq}, \\
  R_{\rm c} &= 8\sqrt{\frac{k_{\rm B}T}{\mu_{\rm SiO_2}}} \frac{a^2}{m H},
\end{align}
{are the evaporation rate and condensation rate, respectively,} where $\mu_{\rm SiO_2}$ is the mean molecular weight for SiO$_2$, $k_{\rm B}$ is the Boltzmann constant, and $P_{\rm eq}$ is the saturated or equilibrium pressure given by the Clausius-Clapeyron equation
\begin{equation}
  P_{\rm eq} = P_{\rm eq,0} e^{-T_a/T},
\end{equation}
where $T_a$ and $P_{\rm eq,0}$ are constants depending on the species.  For SiO$_2$, we use the empirical formula from \citet[][see also \citealt{Misener2022}]{Visscher2013}
\begin{equation}\label{eq:P_eq}
  \log_{10} P_{\rm eq} \left(\mathrm{SiO}_{2}, \text{liq }\right)=8.203-25898.9 / T
\end{equation}
and assume that there is a finite melted liquid boundary layer, where $T_a$ gives a corresponding latent heat $L_0 \simeq T_a R_{\rm g}/\mu_{\rm SiO_2} = 8.26 \times 10^{10}$ erg g$^{-1}$.

To solve the evaporation and condensation for dust with a size distribution, we re-write Equations \ref{eq:dot_Sigma_v} and \ref{eq:dot_Sigma_d} in a discrete, semi-implicit way
\begin{align}
  \Sigma_v^{n+1} - \Sigma_v^{n} &= R_{\rm e} \Sigma_{\rm d}^{n+1} \Delta t - R_{\rm c} \Sigma_{\rm v}^{n+1} \Sigma_{\rm d}^{n} \Delta t, \\
  \Sigma_d^{n+1} - \Sigma_d^{n} &= R_{\rm c} \Sigma_{\rm v}^{n+1} \Sigma_{\rm d}^{n} \Delta t - R_{\rm e} \Sigma_{\rm d}^{n+1} \Delta t,
\end{align}
where $n$ and $n+1$ denote two consecutive time steps and $\Delta t$ is the length of a time step.  We further substitute $\Sigma_{\rm d}$ with $\sum\limits_k \sigma_k \ \textnormal{d}\log a$ from each mass bin and organize the equations into a linear system
\begin{align}
  (1 + R_{\rm c}' \Delta t) \Sigma_{\rm v}^{n+1} - R_{\rm e}\Delta t \sum\limits_k \sigma_k^{n+1} \textnormal{d}\log a &= \Sigma_{\rm v}^{n}, \\
  -R_{\rm c}' \Delta t \Sigma_{\rm v}^{n+1} + (1 + R_{\rm e}\Delta t) \sigma_k^{n+1} \textnormal{d}\log a &= \sigma_k^{n} \textnormal{d}\log a, \label{eq:sigma_k}
\end{align}
where $R_{\rm c}' = R_{\rm c} \sigma_k^{n} \textnormal{d}\log a$.  Equation \ref{eq:sigma_k} represents a series of equations with $k = 1, \cdots, N_k$, where $N_k$ is the total number of dust species in the model.  The evolution of evaporation and condensation can be then solved in the matrix format
\begin{equation}
  \begin{pmatrix}
    1 + R_{\rm c}' \Delta t & -R_{\rm e}\Delta t     & \cdots   & -R_{\rm e}\Delta t    \\
    -R_{\rm c}\Delta t      & 1 + R_{\rm e}\Delta t  &          &                       \\
    \vdots                  &                        & \ddots   &                       \\
    -R_{\rm c}\Delta t      &                        &          & 1 + R_{\rm e}\Delta t
  \end{pmatrix}
  \begin{pmatrix}
    \Sigma_{\rm v}^{n+1}                    \\
    \sigma_1^{n+1} \textnormal{d}\log a     \\
    \vdots                                  \\
    \sigma_{N_k}^{n+1} \textnormal{d}\log a \\
  \end{pmatrix}
  = 
  \begin{pmatrix}
    \Sigma_{\rm v}^{n}                    \\
    \sigma_1^{n} \textnormal{d}\log a     \\
    \vdots                                \\
    \sigma_{N_k}^{n} \textnormal{d}\log a 
  \end{pmatrix}.
\end{equation}

Since the condensation rate $\dot{\Sigma}_{\rm v2d}$ depends on the remaining available solid materials, we impose a tiny surface density floor for dust particles smaller than $0.01$ cm such that vapour can condense back to solids when the temperature drops after a full melt-down.  We adopt this idealized prescription due to the non-uniform nature of nucleation in such a turbulent disc environment and its potential enhancement by non-linear effects within local eddies \citep{Tang2022}.  

Generally speaking, the timescale for thermal evolution is often much shorter than the timescale needed for dust size distribution evolution.  However, our calculations are only computationally feasible with time steps based on the latter.  Appendix \ref{app:tests} demonstrates that our thermal module is numerically robust against these time steps.  Furthermore, particles that gain or lose too much mass may leave the original size bin and shift to another one, which is not taken into account in this work.  However, we carefully examine the net effects of evaporation and condensation in our simulations and find that the consequent relative mass changes in all bins are only of the order of $1\%$ at each time step, much less than the fractional change needed to shift bins ($\sim 32\%$) in our setup.

\begin{figure*}
  \includegraphics[width=0.8\linewidth]{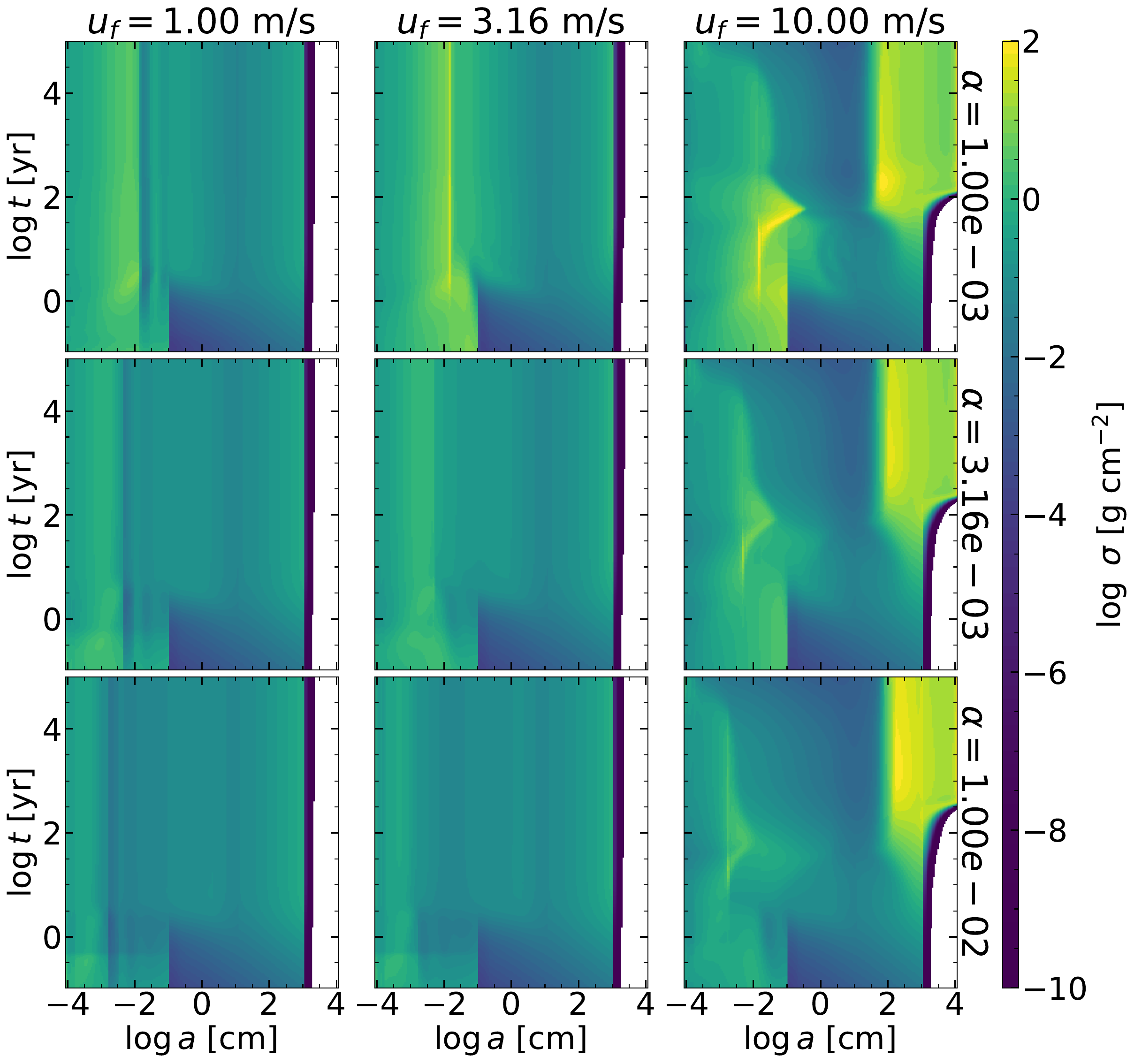}
  \caption{The evolution of vertically integrated dust surface density distribution in the particle size-time plane for all of our full-ingredient dust evolution models, with different $\alpha$ (from \textit{top} to \textit{bottom}) and $u_{\rm f}$ (from \textit{left} to \textit{right}).  In these models, $a_{\rm supp,max}=1000$ cm, $Z_{\rm supp}=0.01$. 
  \label{fig:fiducial_cases}}
\end{figure*}

\begin{figure}
  \includegraphics[width=\linewidth]{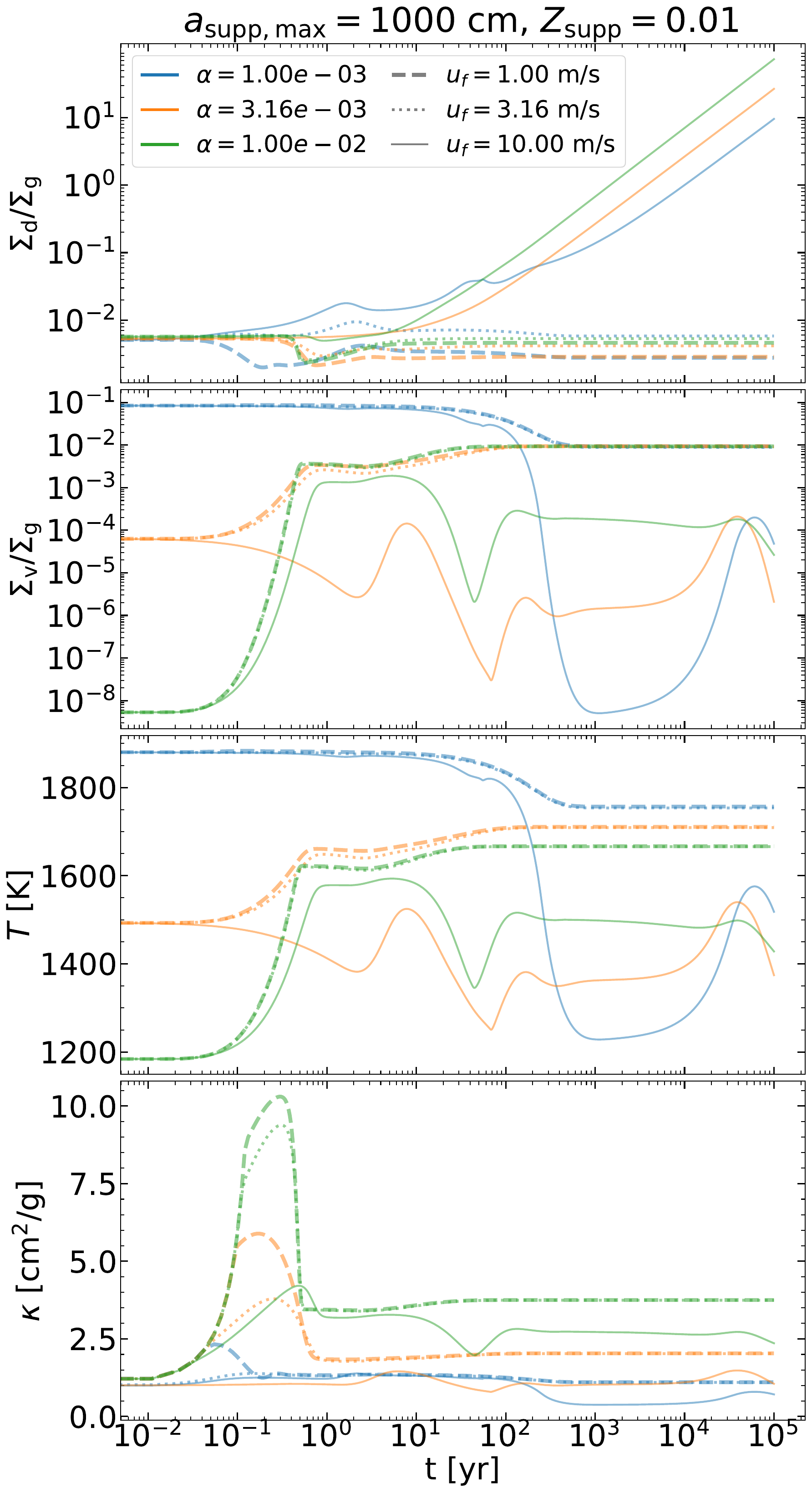}
  \caption{The evolution of (from \textit{top} to \textit{bottom}) dust-to-gas surface density ratio, vapour-to-gas surface density ratio, local disc temperature, and total opacity for the same models shown in Fig. \ref{fig:fiducial_cases}, color-coded by $\alpha$ and line-style-coded by $u_{\rm f}$. 
  \label{fig:fiducial_cases_disc_bulk_prop}}
\end{figure}

\begin{figure}
  \includegraphics[width=\linewidth]{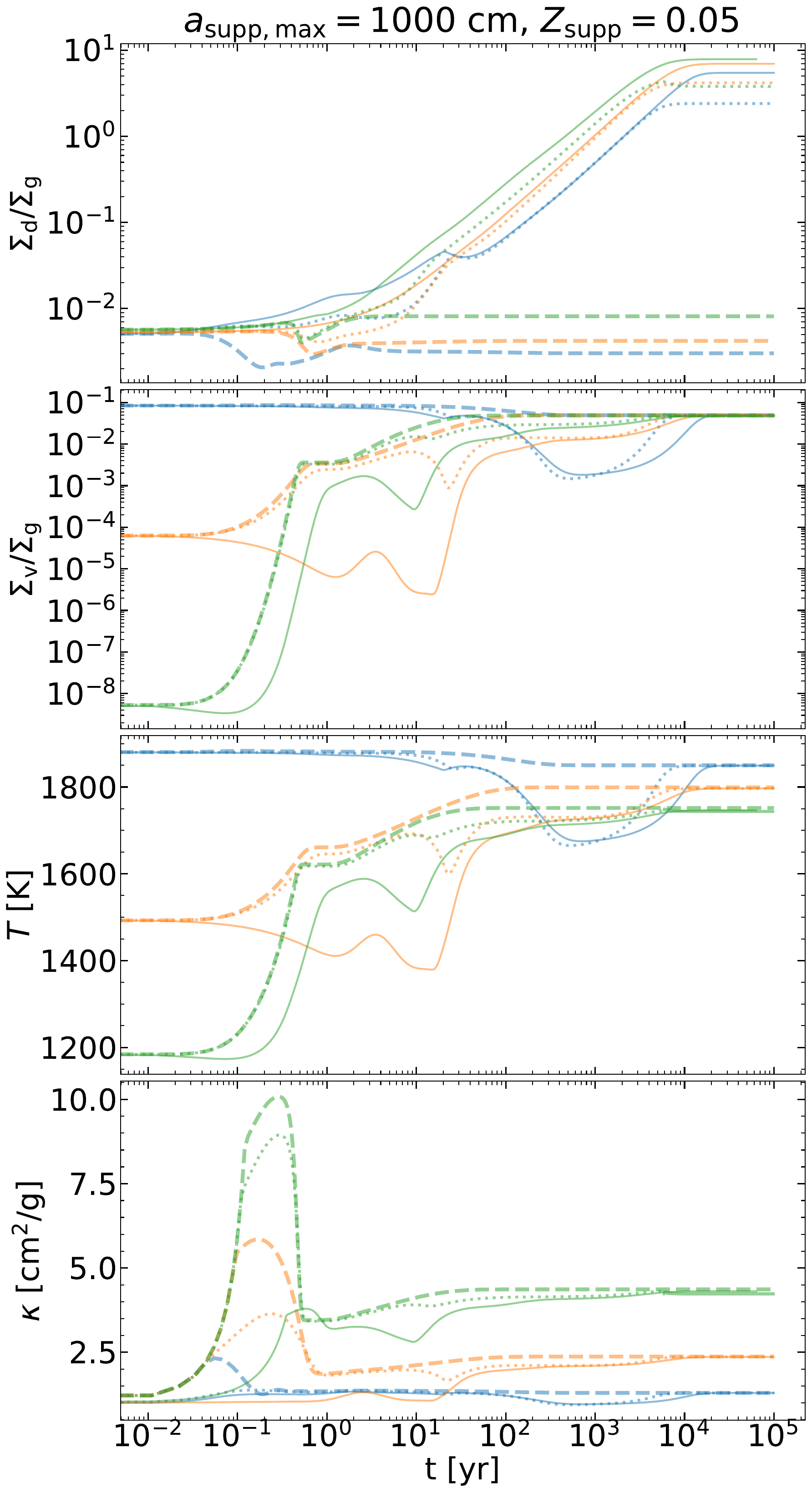}
  \caption{Similar to Fig. \ref{fig:fiducial_cases_disc_bulk_prop} but for models with $Z_{\rm supp}=0.05$.
  \label{fig:fiducial_cases_disc_bulk_prop_Z5}}
\end{figure}

\subsection{Numerical Setup}
\label{subsec:setup}

Table \ref{tab:paras} summarizes the physical and numerical parameters for our dust evolution simulations, where we adopt the same disc model outlined in \citetalias{Li2022} as the initial conditions (see Sections 2 -- 4 therein for detailed descriptions).  Similarly, the dust evolution is then determined by the $\alpha$-viscosity, the fragmentation velocity threshold $u_{\rm f}$, the dust-to-gas ratio of the accreted materials from the external disc $Z_{\rm supp}$, and the maximum supplied particle size $a_{\rm supp,max}$.

In this work, we depart from the conventional approach of initializing all solids with a monodispersed distribution at the micron scale, which results in artificially elevated dust opacity.  Instead, the initial dust size distribution follows the MRN power-law distribution, ranging from $10^{-4}$ to $10^{-2}$ cm.  This choice enables us to maintain self-consistency with our disc model by normalizing the overall dust surface density to be $\lesssim 1\%$ of the gas surface density and achieve a desired initial opacity of $\kappa = 1.0\ {\rm cm}^2/{\rm g}$.  For the MRN distribution, the value of $\kappa$ is relatively insensitive to the initial disc temperature.

The vapour surface density is initialized with the equilibrium value that can be derived from \ref{eq:P_eq} given a certain temperature
\begin{equation}
    \Sigma_{\rm v,0} = 26\times 10^{8.203 - \frac{25898.9}{T_{\rm rad,0}}} \left( \frac{\text{bar}}{\text{g }\text{cm}^{-1} \text{ s}^{-2}} \right) \frac{\sqrt{2\pi} H}{c_{\rm s}^2},
\end{equation}
where $T_{\rm rad,0}$ is the initial temperature at disc midplane and $c_{\rm s}$ denotes gas sound speed.  {During the evolution, we further assume that the accretion funnels remove vapor in the same manner as gas accretion, i.e., $\dot{\Sigma}_{\rm v} \equiv \Sigma_{\rm v}  \dot{\Sigma}_{\rm g} / \Sigma_{\rm g}$.}

Our models take into account all the physical ingredients considered in \citetalias{Li2022} (e.g., mass transfer, feedback effects), besides the newly included dust thermal evolution.  We again only evolve all the models for $10^5$ yr and consider $Z_{\rm final} \gtrsim 1$ as the criterion for significant solid accumulation.

In this work, we focus on the temperature evolution of the gas disc following Equation \ref{eq:energy} and maintain a constant background gas density $\Sigma_{\rm g}$ at the accumulation site for simplicity.  In reality, the gas surface density profile could be altered on the viscous timescale because the inflow of materials tends to smooth out discontinuities in the disc accretion rate $\dot{M}$, which is directly proportional to $\Sigma_{\rm g} T$.  However, this alteration occurs on a timescale significantly longer than the thermal evolution and so we simplify our model by neglecting the time evolution of $\dot{M}$ and $\Sigma_{\rm g}$.  Furthermore, the assumption of a constant $\Sigma_{\rm g}$ breaks down in the limit where the vapour surface density $\Sigma_{\rm v}$ is a significant fraction of $\Sigma_{\rm g}$, which however is never the case in all of our results.  

\begin{figure*}
  \centering
    \makebox[\textwidth][c]{ 
    \includegraphics[width=1.\linewidth]{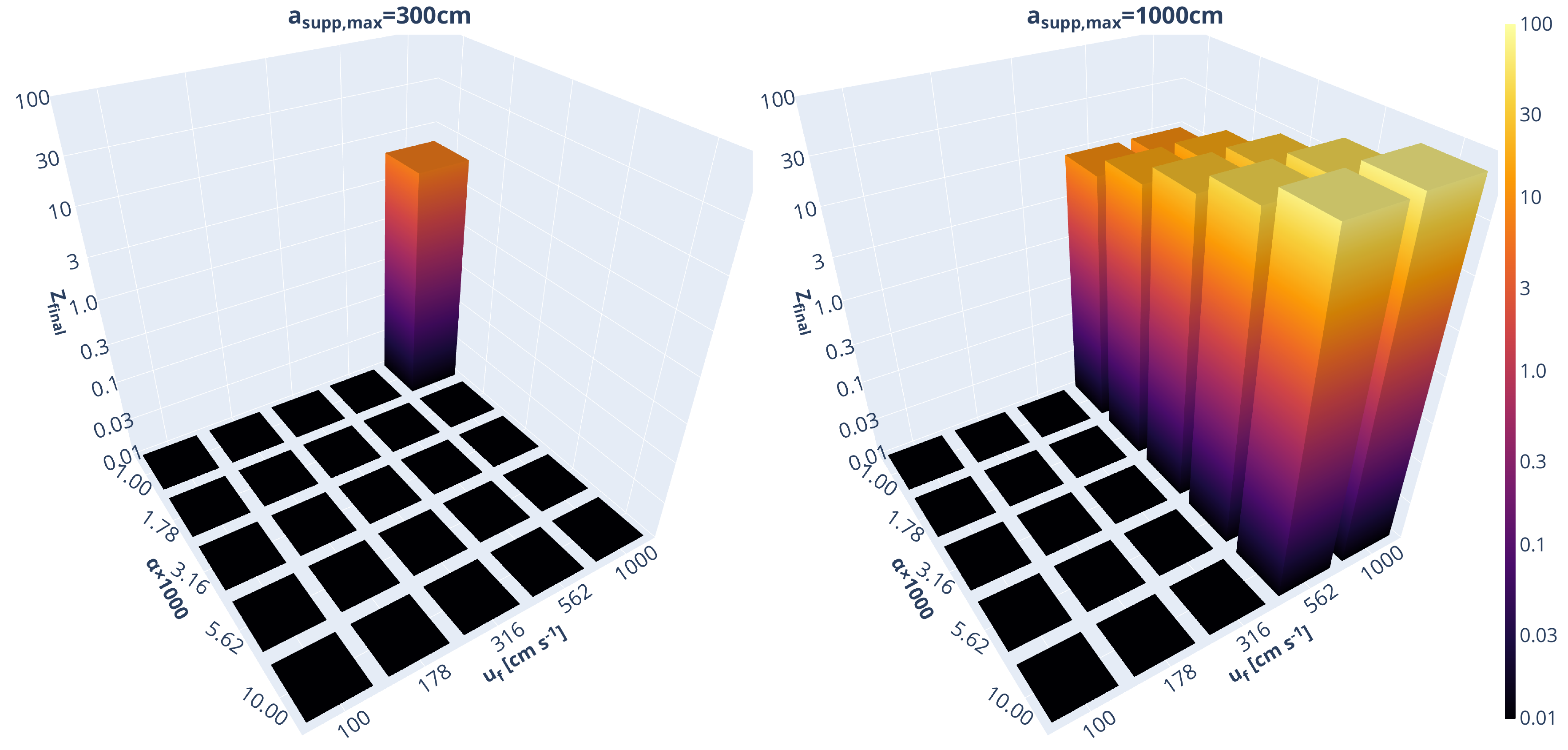}
    }
  \caption{Final dust-to-gas surface density ratios $Z_{\rm final}$ after $10^5$ years of local dust evolution for our full-ingredient dust evolution models with $Z_{\rm supp}=0.01$.  Models with $a_{\rm supp,max} = 300$ cm (\textit{left}) do not accumulate considerable dust mass and produce almost identical results.  Models with $a_{\rm supp,max} = 1000$ cm (\textit{right}) may yield runaway growth and significant solid accumulation given ideal disc conditions.   An interactive version of this plot is available at \href{https://rixinli.com/RubbleSurveyResultsII.html}{https://rixinli.com/RubbleSurveyResultsII.html}
  \label{fig:Zfinal_3D_bars}}
\end{figure*}

\begin{figure*}
  \centering
    \makebox[\textwidth][c]{ 
    \includegraphics[width=1.\linewidth]{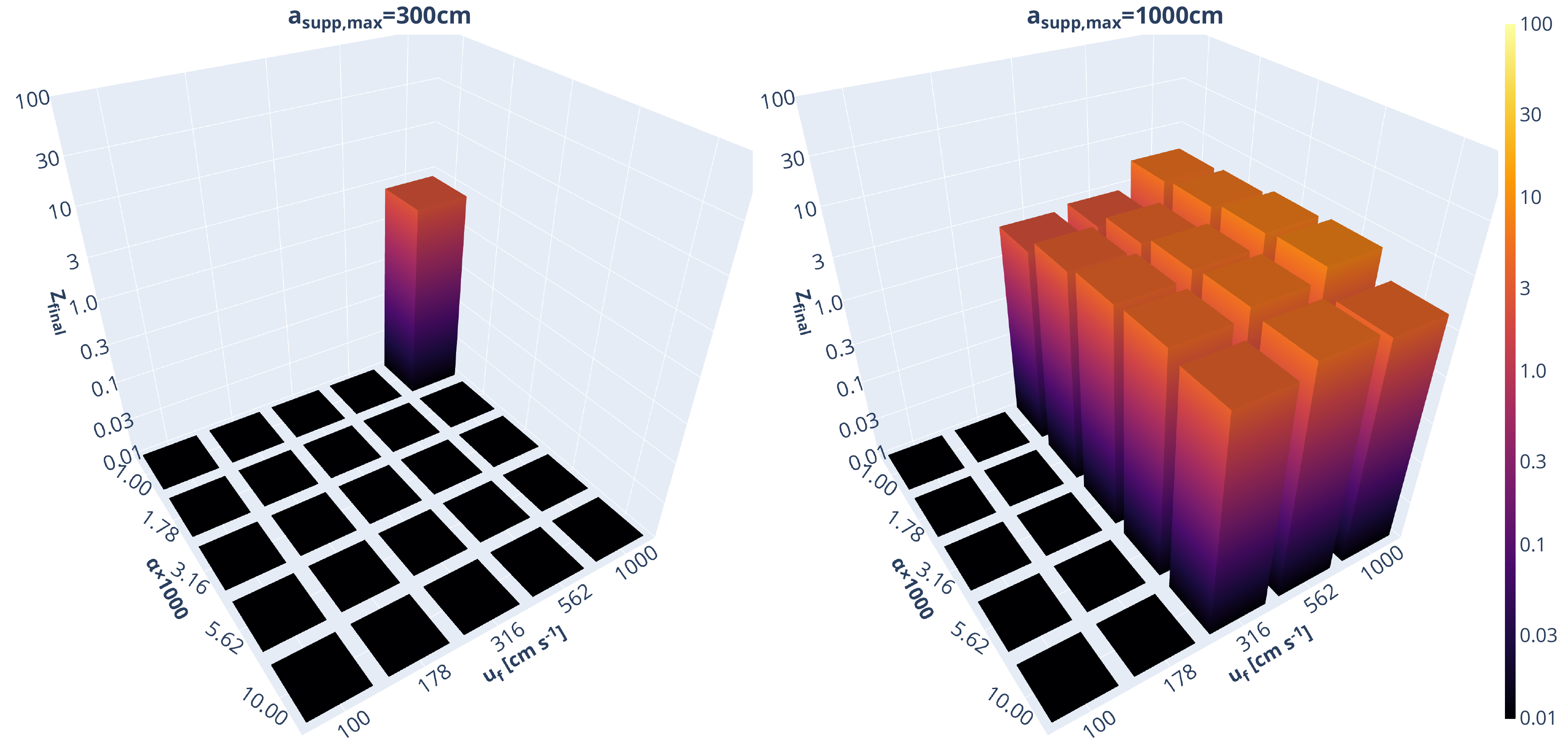}
    }
  \caption{Similar to Figure \ref{fig:Zfinal_3D_bars} but for models with $Z_{\rm supp}=0.05$.  An interactive version of this plot is available at \href{https://rixinli.com/RubbleSurveyResultsII.html}{https://rixinli.com/RubbleSurveyResultsII.html}
  \label{fig:Zfinal_3D_bars_Z5}}
\end{figure*}

\section{Results}
\label{sec:results}

We conduct a suite of local dust evolution simulations, incorporating self-consistent opacity-driven disc thermal evolution, with parameters listed in Table \ref{tab:paras}.  Section \ref{subsec:fiducial_model} first presents a set of representative cases and elucidate the interplay between the collisional and thermal dust evolution.  Section \ref{subsec:survey_results} then surveys disc conditions that are required to achieve effective dust retention.

\subsection{Fiducial Model}
\label{subsec:fiducial_model}

Figs. \ref{fig:fiducial_cases} and \ref{fig:fiducial_cases_disc_bulk_prop} compare the time evolution of dust size distribution, the dust/vapour-to-gas ratio, the gas disc temperature, and the total disc opacity between our models with different $\alpha$ and $u_{\rm f}$, where $a_{\rm supp,max}$ is fixed to $1000$ cm and $Z_{\rm supp}$ is fixed to $0.01$.

We find that almost all models experience significant thermal evolution in the very first year, before reaching a steady state or when the external dust supply makes a difference.  For cases with a higher $\alpha$ and thus a relatively lower initial disc temperature $T_{\rm rad,0}$, dust fragments easily and quickly increases the total opacity $\kappa$ and disc temperature $T$, unless $u_{\rm f}$ is high enough.  Conversely, in cases with a lower $\alpha$ (i.e., higher $T_{\rm rad,0}$), dust particles may melt down faster than fragmentation such that the disc temperature does not change substantially, particularly with a lower $u_{\rm f}$ where smaller dust dominates.

Efficient breakthrough and sweep-up growth happen in models with a larger $u_{\rm f}$, leading to significant dust accumulation.  In these cases, the higher fragmentation velocity threshold promotes coagulation and allows the largest supplied particles to pile up, which eventually results in runaway growth and breakthrough.  Once there are enough large particles ($> 10^{3.5}$ cm), sweep-up growth via mass transfer dominates the dust evolution and transports almost all the supplied dust mass into very large particles \citep{Windmark2012b}.

Although similar dust retention processes have been seen in \citetalias{Li2022}, their dynamic impact to the disc thermal evolution is shown for the first time in this work.  We find that the initial pile up of solids near $a_{\rm supp,max}$ (at $t\sim 20$ yr) cause a deficit of sub-mm dust and consequently a decrease in the total opacity and a sharp drop in the disc temperature.  However, a considerably lower disc temperature hinders dust loss due to fragmentation (and funnel flows) and inflates the dust surface density across the entire size distribution, which then in response leads to an upturn for $\kappa$ and $T$.  After breakthrough happens and most of the solid mass resides in the very large particles, the changes in $\kappa$ and $T$ become subtle.
Furthermore, the final disc temperature always shows a much weaker dependency on $\alpha$ in comparison to the initial temperature, regardless of $Z_{\rm final}$ and accompanied by different extent of offset in the opacity from the initial value of unity.

An accelerated external disc dust supply augments the overall reservoir of solid mass, which always facilitates dust accumulation as seen in \citetalias{Li2022}. However, when thermal evolution is taken into consideration, this increased supply also elevates the abundance of fine dust particles that raise disc opacity.  As illustrated in Figure \ref{fig:fiducial_cases_disc_bulk_prop}, models with $Z_{\rm supp}=0.05$ consistently achieve a higher opacity and thus a higher disc temperature when compared to their counterparts in Fig. \ref{fig:fiducial_cases_disc_bulk_prop}.  Therefore, although more cases can generate substantial dust accumulation, the dust-to-gas ratios eventually converge to a plateau after $\sim 10^{4}$ yr, representing a new balance between the thermal evolution and the dust size distribution evolution.

\subsection{Threshold for significant dust accumulation}
\label{subsec:survey_results}

Figs. \ref{fig:Zfinal_3D_bars} and \ref{fig:Zfinal_3D_bars_Z5} present our survey results on parameters listed in Table \ref{tab:paras} and show which runs retain substantial dust mass and which do not via the final dust-to-gas ratio $Z_{\rm final}$.  We find that the maximum-supplied size $a_{\rm supp,max}$ and the fragmentation velocity threshold $u_{\rm f}$ largely determine the dust evolution, consistent with our findings in the last section.  When $a_{\rm supp,max} = 1000$ cm, substantial dust accumulation happens when $u_{\rm f} \gtrsim 5.62$ m s$^{-1}$ if $Z_{\rm supp} = 0.01$, or $u_{\rm f} \gtrsim 3.16$ m s$^{-1}$ if $Z_{\rm supp} = 0.05$.  When $a_{\rm supp,max} \lesssim 300$, only the corner case ($\alpha=10^{-3}$, $u_{\rm f}=10$ m s$^{-1}$) can efficiently retain dust.

As discussed in the previous section, for $a_{\rm supp,max} = 1000$ cm, the number of cases with $Z_{\rm final} \gtrsim 1$ increases with $Z_{\rm supp}$.  However, the values of $Z_{\rm final}$ in these runs with $Z_{\rm supp}=0.05$ are consistently lower than those in their counterparts with $Z_{\rm supp}=0.01$, due to the higher opacity/temperature caused by the richer supply of fine dust grains.  Therefore, with thermal evolution taken into account, a larger dust supply may not always yield a positive impact on all aspects of the dust accumulation process.  It is possible that a even higher $Z_{\rm supp}$ could fully suppress dust retention, which is however beyond the scope of this work given that the choice of $Z_{\rm supp}=0.05$ is already quite large. 

Compared to the previous results that neglected thermal evolution in \citetalias{Li2022}, significant accumulation of mm--cm dust through the \textit{feedback+GI scenario} (i.e., most effective at the low $\alpha$ high $u_{\rm f}$ parameter space) becomes unviable, as small dust particles contribute most to the opacity/temperature enhancement, which in turn quench dust retention.  Instead, the \textit{breakthrough scenario} exhibits a better chance of accumulating solids, with a strong preference for a higher $u_{\rm f}$ and a moderate preference for a higher $\alpha$, which provides a lower initial disc temperature.

\begin{figure*}
  \includegraphics[width=\linewidth]{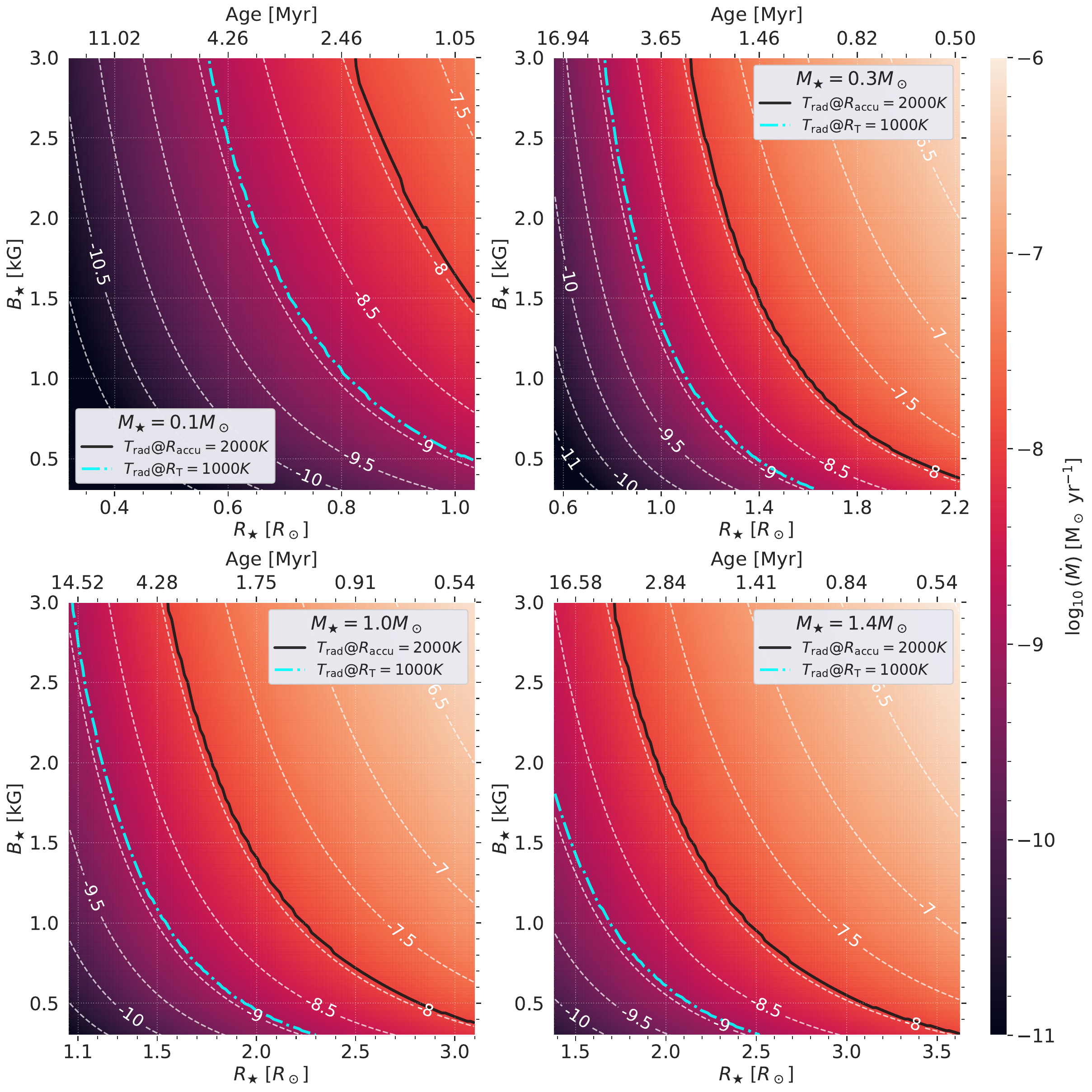}
  \caption{The accretion rate in the stellar radius vs. stellar magnetic field strength frame for different stellar mass (from \textit{left/top} to \textit{right/bottom}: $0.1M_\odot$, $0.3M_\odot$, $1.0M_\odot$, $1.4M_\odot$), assuming that the magnetospheric truncation radius converges with the corotation radius, $P_\bigstar=8$ days, $f_{\rm out}=1.25$, and $\alpha=0.01$.  The top axis represents the stellar age corresponding to the stellar radius on the bottom axis, as determined by the stellar evolution model in \citet{Baraffe2015}.  The \textit{black solid} curve and the \textit{blue dash dotted} curve sandwich the area where the disc temperature $T_{\rm rad}$ at $R_{\rm T}$ is above $10^3$ K (such that disc is truncated effectively) and $T_{\rm rad}$ at $R_{\rm accu}$ is below $2000$ K (such that refractory dust may survive), allowing potential dust trapping and retention.
  \label{fig:Mdot_map_alpha1e-2}}
\end{figure*}

\begin{figure*}
  \includegraphics[width=\linewidth]{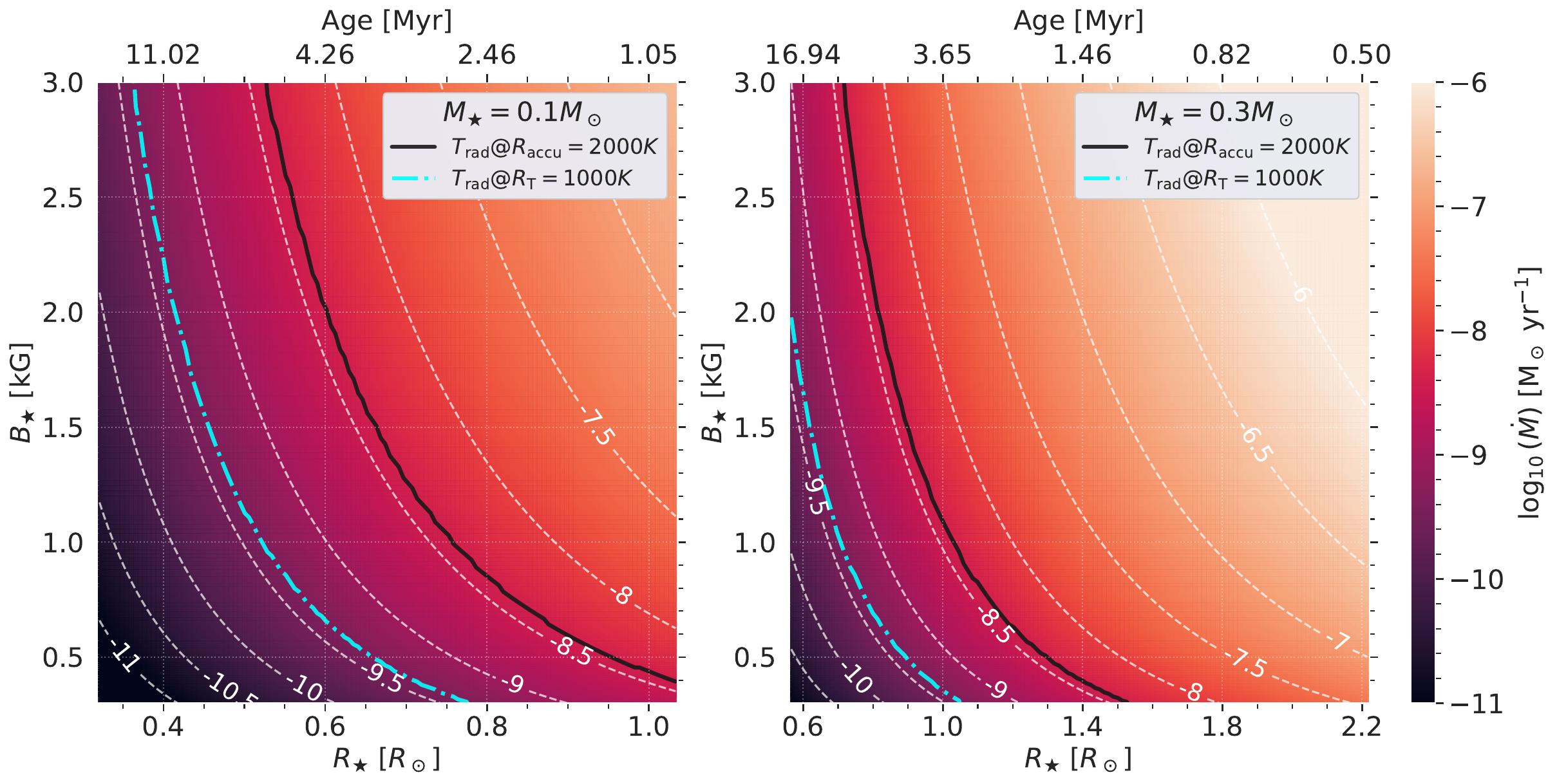}
  \caption{Similar to the top row of Fig. \ref{fig:Mdot_map_alpha1e-2} but with the assumption of $P_\bigstar=4$ days.
  \label{fig:Mdot_map_alpha1e-2_4d}}
\end{figure*}

\section{Summary and Discussions}
\label{sec:final}

In this sequel to our prior work in \citetalias{Li2022}, we delve into the intricate dynamic synergy between the local dust distribution evolution and the local disc thermal evolution near the truncated inner boundary of protoplanetary discs.  In particular, we improve our model by taking into account the self-consistent intertwined changes, including the change in disc temperature due to the change in opacity, the change in opacity due to the change in dust size distribution, and the evaporation and condensation of solids at all sizes due to the evolving disc temperature.

Our main findings are as follows:
\begin{itemize}
    \item We find that the dust thermodynamics plays an important role in dust size evolution.  Since the total opacity is sensitive to the amount of small ($\sim$ mm) dust particles, 
    the self-consistent thermal evolution fully suppresses the \textit{feedback+GI scenario} that builds a mass reservoir in small dust particles until the dust-to-gas ratio exceeds order unity. 
    Thus, significant dust accumulation only happens in the \textit{breakthrough scenario} that overcomes the fragmentation barrier and builds a mass reservoir in the runaway growth of large rocky particles. 
    Furthermore, the final opacity is self-regulated at the order of a few in all cases, 
    roughly consistent with the assumed unity opacity adopted in the initial radiative disc model.
    \item A successful \textit{breakthrough scenario} favours a larger fragmentation velocity threshold and larger particles in the accreted materials from the external disc, 
    which makes breakthrough growth easier.  However, a high dust-to-gas ratio in the supplied materials may hinder dust retention as it takes a longer time to process small dust particles in the size distribution evolution, leading to a higher opacity and therefore a hostile higher disc temperature.
\end{itemize}

\citet{Kunimoto2020} suggested that the occurrence rate of close-in Kepler planets generally increases from early-type to late-type around FGK stars.  To explore how the dust evolution scenario proposed in this work varies around stars with different masses, Fig. \ref{fig:Mdot_map_alpha1e-2} shows the range of the accretion rate $\dot{M}$ that meets the temperature requirements of our scenario, that is, the initial disc midplane temperature $T_{\rm rad,0}$ at the truncation radius $R_{\rm T}$ is above $1000$ K (such that disc is truncated) and $T_{\rm rad,0}$ at the dust accumulation radius $R_{\rm accu}$ is below $2000$ K (such that dust can survive from sublimation), as a function of stellar properties $R_\star$, $B_\star$, and $M_\star$ (radius, strength of dipole magnetic field, and mass; similar to Fig. 2 in \citetalias{Li2022}).

Under the assumptions of a fixed stellar spin period ($P_\star = 8$ days) and that the corotation radius $R_{\rm Co}$ converges with the truncation radius $R_{\rm T}$, we find that the viable accretion rate always 
\footnote{The reason for the persistent range of $\dot{M}$ is that the disc midplane temperature $T_{\rm rad} \propto r_{\rm au}^{-0.9} m_\star^{0.3} \propto m_\star^{0}$, where $r_{\rm au} = R/(1 \text{au}) = R_{\rm T} \simeq R_{\rm Co} \propto m_\star^{1/3}$.}
falls between $\sim 10^{-9}$ -- $10^{-8} M_\odot$ yr$^{-1}$.  However, many uncertainties remain.  For example, $P_\star$ not only evolves with stellar evolution but also exhibits a very broad distribution, particularly for low-mass stars \citep[see Fig. 4 in][]{Reiners2022}.  Moreover, $B_\star$ decreases with $P_\star$, but with a large scatter within each stellar mass interval.  These uncertainties make it hard to pin down the range of stellar age -- inferred from $R_\star$ based on the stellar evolution model in \citet{Baraffe2015} -- for the proposed dust retention scenario to happen. 

Taking $P_\star=4$ days as an example, Fig. \ref{fig:Mdot_map_alpha1e-2_4d} presents the viable accretion rate range for low-mass stars that are rapid rotators.  Compared to the slow rotators shown previously in Fig. \ref{fig:Mdot_map_alpha1e-2}, these rapid ones will need to evolve to an even later stage to set off dust accumulation as the viable range of $\dot{M}$ decreases to $\sim 10^{-9.5}$ -- $10^{-8.5} M_\odot$ yr$^{-1}$.  This comparison suggests that our proposed scenario for dust retention and potentially subsequent planet formation likely favours slower rotators, which is somewhat more abundant at lower stellar masses \citep{Reiners2022, Stefansson2023}, roughly consistent with the occurrence rate findings \citep{Kunimoto2020}.

\section*{Acknowledgements}

We thank Gibor Basri, Lee Hartmann, Gregory Herczeg, and Caeley Pittman for useful conversation.
R.L. acknowledges support from the Heising-Simons Foundation 51 Pegasi b Fellowship.

\section*{Data Availability}

The data in this article are available from the corresponding author on reasonable request. 



\bibliographystyle{mnras}
\bibliography{refs}



\appendix
\section{Numerical Tests on Thermodynamics}
\label{app:tests}

In this section, we test the numerical robustness of the evaporation and condensation module implemented in our code.  Particularly, since the timescale for thermal evolution is often much smaller than the timescale for dust size evolution, we do not evolve the dust size distribution and only compute the thermal evolution under a fixed disc temperature $T_0$ against a vast range of time steps, ranging from $10^{-2}$ to $10^{4}$ seconds.  

In all of our tests, we initialize the dust size distribution with an MRN power-law distribution, ranging from $10^{-4}$ and $10^{-2}$ cm, with a total dust surface density of $\Sigma_{\rm d,0}=100$ g cm$^{-1}$.  The initial vapour surface density is $\Sigma_{\rm v,0}=1$ g cm$^{-1}$.  Under a given temperature, the equilibrium vapour surface density can be derived from Eq. \ref{eq:P_eq} as
\begin{equation}
    \Sigma_{\rm v,eq} = 26\times 10^{8.203 - \frac{25898.9}{T_0}} \left( \frac{\text{bar}}{\text{g }\text{cm}^{-1} \text{ s}^{-2}} \right) \frac{\sqrt{2\pi} H}{c_{\rm s}^2},
\end{equation}
where $H$ and $c_{\rm s}$ are the gas scale height and sound speed. 

\begin{figure*}
  \includegraphics[width=0.48 \linewidth]{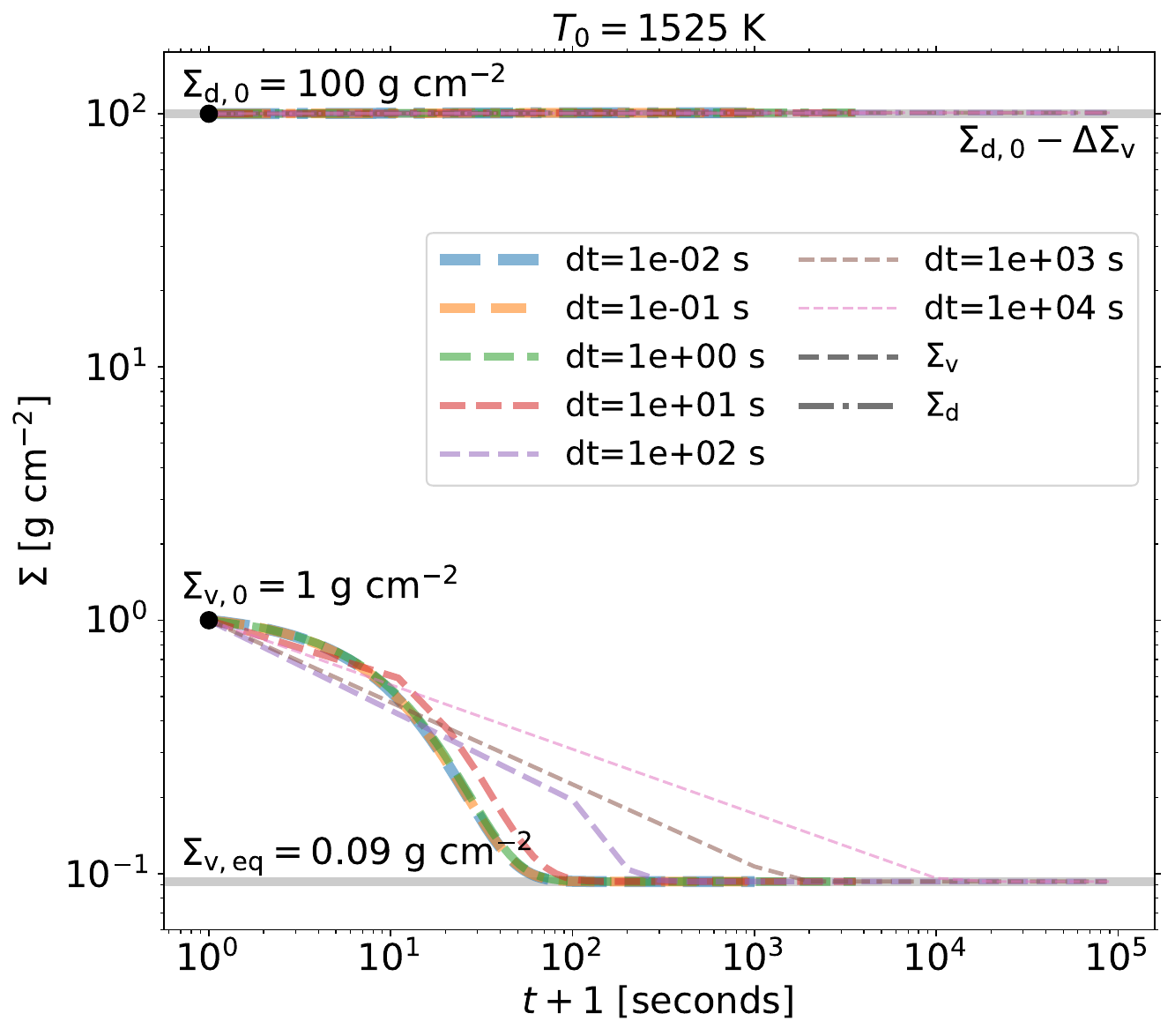}
  \includegraphics[width=0.48 \linewidth]{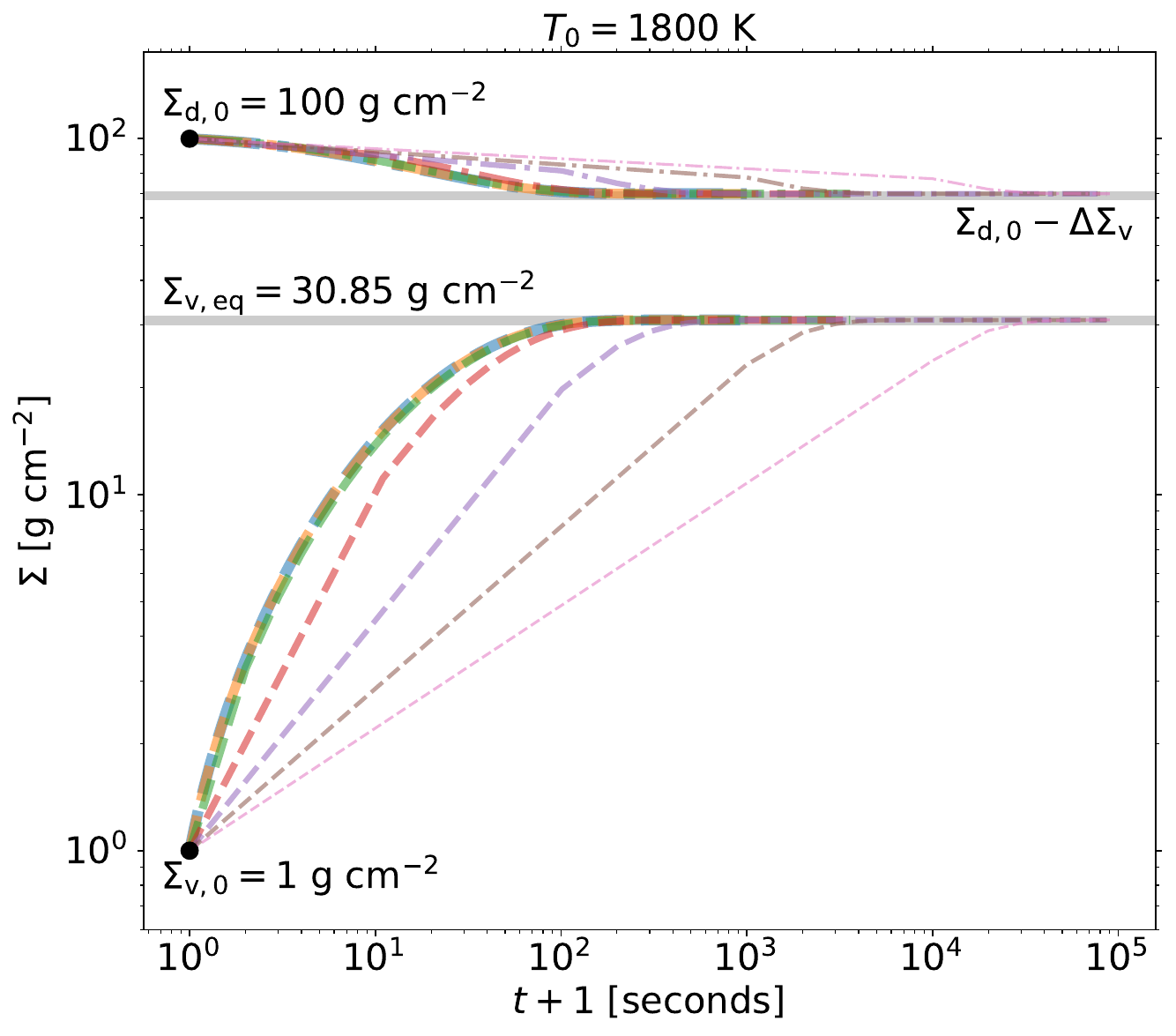}
  \includegraphics[width=0.65 \linewidth]{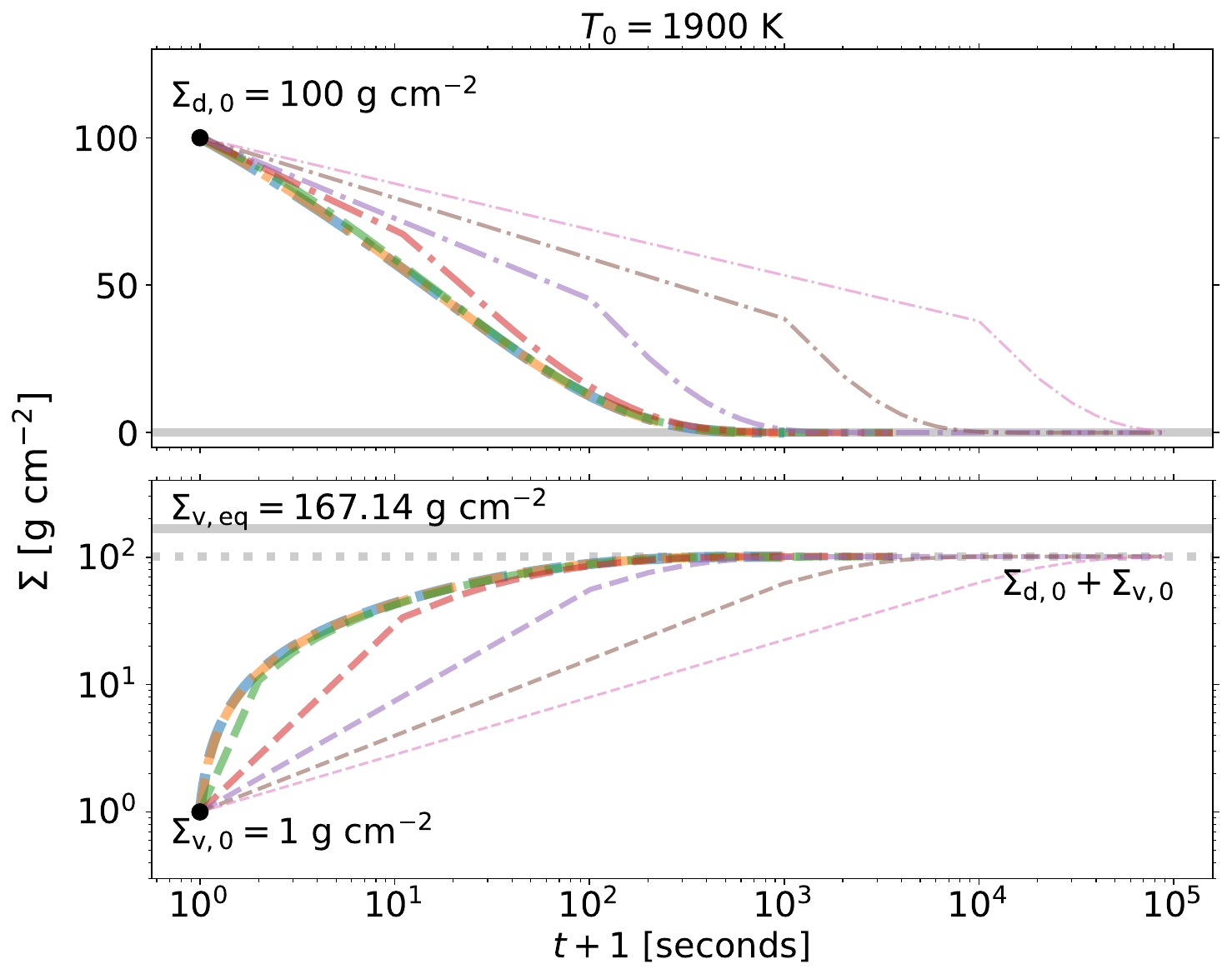}
  \caption{The evolution of dust surface density ({\it dashed lines}) and vapour surface density ({\it dash-dotted lines}) with a fixed initial dust size distribution, under a fixed temperature $T_0$, and against a range of time steps ({\it colour-coded} and {\it in a decreasing line width}).  
  Both $\Sigma_{\rm d}$ and $\Sigma_{\rm v}$ quickly evolves to the expected equilibrium values ({\it grey solid/dotted lines}) regardless of the time step length.
  \label{fig:ce_test1}}
\end{figure*}

Fig. \ref{fig:ce_test1} shows that the evolution of the surface densities of dust and vapour under three different temperatures: $T_0=1525$ K, $1800$ K, and $1900$ K.  Both $\Sigma_{\rm d}$ and $\Sigma_{\rm v}$ quickly evolve to the expected equilibrium values, regardless of the time step length or the major direction of the thermal evolution (either evaporation or condensation).  In the case with $T_0 = 1900 K$, since $\Sigma_{\rm v,eq}$ is larger than the available mass budget, all solids become vapourized.

We have also conducted extra test simulations with different fixed dust size distribution, including monodispersed dust distributions at different sizes, random two-species distributions, or quasi-equilibrium dust distributions in interplanetary discs (\citealt{Birnstiel2011}; see also Appendix A2 in \citetalias{Li2022}), all of which produce similar results.  Our comprehensive tests demonstrate that the semi-implicit scheme described in Section \ref{subsec:ec} produces consistent and accurate results, even at a large time steps that are used when computing dust size distribution evolution.

\bsp    
\label{lastpage}

\end{document}